\documentclass[11pt,a4paper]{article}
\usepackage{jcappub}

\usepackage[latin9]{inputenc}
\usepackage{units}
\usepackage{amssymb}
\usepackage{setspace}
\usepackage{esint}


\title{The creation of radiation and the relic of inflaton potential}

\author{Yu-chung Chen}

\affiliation{Department of Physics, National Taiwan University,}

\affiliation{Leung Center for Cosmology and Particle Astrophysics, National Taiwan University,}

\affiliation{Taipei 10617, Taiwan, R.O.C.}

\emailAdd{yuchung.chen@gmail.com}

\abstract{
Recently, we have performed research on the subject of the cosmological
constant problem. The scenario is based on two postulates for inflationary
theory: one is that inflaton $\phi$ can interact with radiation (relativistic
particles); the other is that radiation will be created continuously
during and after the epoch of inflation. According to these postulates
and from a \textquotedblleft{}macroscopic perspective\textquotedblright{},
we discover that radiation can be viewed as a product of the interaction
between $\dot{\phi}$ and some ``effective kinetic frictional force''
that exists in inflaton dynamics. Deducing and surmising from \textquotedblleft{}effective
friction\textquotedblright{}, we obtain conclusions of two special
types of expanding universe: A Type I universe will finally enter
an expanding course after a special time $t_{*}$ with uniformly rolling
$\dot{\phi}\left(t_{*}\right)$ due to the balance between $V^{'}\left(\phi\left(t\right)\right)$,
$3H\left(t\right)\dot{\phi}\left(t_{*}\right)$ and the \textquotedblleft{}effective
kinetic frictional force\textquotedblright{}. In this result, the
expanding course will see particles created continuously. Additionally,
for a Type II universe, $\phi$ will be at rest after $t_{r}$ inside
a region named the \textquotedblleft{}stagnant zone\textquotedblright{}
that is formed by the \textquotedblleft{}maximum effective static
frictional force\textquotedblright{}. Consistent with this, inflaton
potential will survive as a relic $V\left(\phi\left(t_{r}\right)\right)$,
playing the role of the effective cosmological constant $\Lambda$.
}


\dedicated{
Dedicated to my beloved daughter CoCo.
}

\begin{document}
\maketitle
\flushbottom

\section{Introduction} \label{sec:1}

According to the research and observations of Friedman \cite{Friedman},
Lema{\^i}tre \cite{Lemaitre} and Hubble \cite{Hubble}, Einstein
and other physicists have been told that it is not necessary to insert
a cosmological term $\lambda$ into the general theory of relativity%
\footnote{Based on his belief in Mach's principle, in 1917 Einstein \cite{Einstein}
inserted the cosmological term $\lambda$ into his new theory of gravity,
as $R_{\mu\nu}-\lambda g_{\mu\nu}=-\kappa\left(T_{\mu\nu}-\frac{1}{2}g_{\mu\nu}T\right)$,
to keep the universe static. In this equation, the coefficient of
general relativity is $\kappa=\nicefrac{8\pi G}{c^{4}}$; $R_{\mu\nu}$
is the Ricci tensor; $g_{\mu\nu}$ is the metric tensor; $T_{\mu\nu}$
is the energy-momentum tensor; and $T=g^{\alpha\beta}T_{\alpha\beta}$.%
}. Therefore, until 1998 most believed that the expansion of our universe
is --- or will be --- slowing down. Nevertheless, much observational
data from \cite{obs(SN)-01,obs(SN)-02,obs(SN)-03,obs(SN)-04,obs(SN)-05,obs(SN)-06,obs(XC)}
provides evidence to oppose intuition: our universe is presently expanding
with acceleration. Besides, \cite{obs(CMB)-01,obs(CMB)-02,obs(CMB)-03,obs(BAO),obs(WGL)-01,obs(WGL)-02,obs(WGL)-03,obs(WGL)-04,obs(WGL)-05,obs(WGL)-06,obs(WGL)-07}
have been/are being performed to excavate greater understandings:
we now know that the major components required to build the current
universe are roughly 0.008\% radiation, 5\% observable matter and
22\% dark matter. In particular, the surplus energy density that we
call dark energy confirms that a new discovery, i.e. accelerating
expansion, is needed.

Obviously, the part of dark energy is a compelling mystery because
few of its properties are known. Those which we do have knowledge
of are as follows: to begin, the first Friedman equation without the
cosmological term,
\begin{equation}
\frac{\ddot{\mathbb{R}}\left(t\right)}{\mathbb{R}\left(t\right)}=-\frac{\kappa c^{2}}{6}\left(\varepsilon+3p\right),\label{eq:1.1}
\end{equation}
shows that the equation of state for dark energy $w_{\mathrm{DE}}\equiv\nicefrac{p}{\varepsilon}<-\nicefrac{1}{3}$
(where $c$ is the speed of light; $\varepsilon$ is the dark energy
density; $p$ is the pressure of dark energy; $\mathbb{R}\left(t\right)$
is the spatial scale factor; and $t$ is cosmic time) should be satisfied
in order to make the \textquotedblleft{}anti-gravity\textquotedblright{},
$\ddot{\mathbb{R}}=\frac{d^{2}\mathbb{R}}{dt^{2}}>0$, become possible
on the large scales of our universe; secondly, the repulsive property
of dark energy requires its distribution to be highly homogeneous and
isotropic; and finally there is still no evidence to suggest that
dark energy interacts with matter through any of the fundamental forces
other than gravity. In order to explain/illustrate this impalpable
phenomenon, many models \cite{mod-01,mod-02,mod-03,mod-04,mod-05,mod-06,mod-07,mod-08,mod-09,mod-10,mod-11,mod-12,mod-13,mod-14,mod-15,mod-16,mod-17,mod(IQ)-01,mod(IQ)-02,mod(IQ)-03,mod(IQ)-04}
have been proposed. Of course, observations also lead to a rethink
of the cosmological term%
\footnote{If we consider that our universe is flat, \cite{obs(CMB)-03} found
that the equation of state for dark energy is $w_{\mathrm{DE}}\simeq-1.08\pm0.12$,
according to the observations of WMAP, SDSS, 2dFGRS and SN Ia. Similarly,
the combined data of BAO, CMB and SNe provides $w_{\mathrm{DE}}\simeq-0.9725\pm0.1255$
\cite{anal}. From this, we cannot abandon the cosmological constant
because, in theory, the equation of state for it $w_{\Lambda}$ is
equal to $-1$.%
} that was thrown in Einstein's trashcan. Picking up the term is excellent
and simple for analyzing the new discovery. Consequently, however,
a question cannot be avoided: What, practically, is the cosmological
term/constant?

To answer, it was thought by \cite{V&CCP-01} that the vacuum energy
density discovered in research on quantum field theory might be the
cosmological constant%
\footnote{Under the terms of the Lorentz invariance for a vacuum energy-momentum
tensor, vacuum energy density $\left\langle \varepsilon_{\mathrm{vac}}\right\rangle $
acts exactly like a cosmological constant since $T_{\mu\nu}^{\mathrm{(vac)}}=\left\langle \varepsilon_{\mathrm{vac}}\right\rangle g_{\mu\nu}$.%
}. Therefore, the Planck vacuum energy density (PVED) could be calculated
by summing the zero-point energies of all normal modes $k$ of some
field of mass $m$ up to a wave cut-off, $k_{\mathrm{cut}}=\sqrt{\pi^{2}c^{3}\left(\hbar G\right)^{-1}}\gg mc\hbar^{-1}$,
as 
\begin{equation}
\left\langle \varepsilon_{\mathrm{Planck}}\right\rangle _{\mathrm{vac}}=\frac{\hbar}{2\left(2\pi\right)^{3}}\int_{0}^{k_{\mathrm{cut}}}4\pi k^{2}dk\sqrt{k^{2}c^{2}+m^{2}c^{4}\hbar^{-2}}\simeq\unitfrac[5.6\times10^{126}]{eV}{cm^{3}}\label{eq:1.2}
\end{equation}
This is provided by the assumption that the smallest limit of general
relativity is the Planck scale. Correspondingly, assuming that the
cosmological constant $\Lambda$ is, in fact, dark energy, its value
is 
\begin{equation}
\varepsilon_{\Lambda}\simeq\frac{3H_{\mathrm{now}}^{2}}{\kappa c^{2}}\times73\%\simeq\unitfrac[3.76\times10^{3}]{eV}{cm^{3}},\label{eq:1.3}
\end{equation}
which can be shown by the Hubble rate at its present day value of
$H_{\mathrm{now}}\approx\unitfrac[70]{\unitfrac{km}{s}}{Mpc}$. Unfortunately,
the PVED is too massive in comparison with the effective density as
witnessed in reality. Other conditions of vacuum energy density, such
as spontaneous symmetry breaking (SSB) in electroweak (EW) theory
and the vacuum transition of quantum chromodynamics (QCD), are also
expelled as candidates for the cosmological constant because we receive
even larger values as 
\begin{equation}
\begin{array}{l}
\left\Vert \left\langle \varepsilon_{\mathrm{EW}}\right\rangle _{\mathrm{SSB}}\right\Vert \thickapprox\unitfrac[10^{56}]{eV}{cm^{3}},\\
\left\langle \varepsilon_{\mathrm{QCD}}\right\rangle _{\mathrm{vac}}\thickapprox\unitfrac[10^{44}]{eV}{cm^{3}}.
\end{array}\label{eq:1.4}
\end{equation}
Two serious problems are yet indicated by \cite{V&CCP-04,V&CCP-04-1}:
(i) Where are these densities? (ii) Why is the cosmological constant
so small?

Even though vacuum energy densities cannot be the cosmological constant,
they are still needed for the investigation of the very early universe.
Realizing the inexplicable problems%
\footnote{These problems are the homogeneous, isotropic, horizon, flatness, initial
perturbation, magnetic monopole, total mass, total entropy and so
on \cite{INF-01}.%
} which emerge when observations are made using Hot Big Bang theory,
Starobinsky, Guth and others devised a beautiful solution: they suggested
that our universe must have inflated itself from a very small size
--- perhaps just a little bigger than a Planck point \cite{INF-02,INF-03,INF-04,INF-05}.
Therefore, according to inflationary theory, a vacuum energy density
dependent on the initial size of the universe is required to trigger
inflation at the beginning. However, a timely mechanism is also needed
to cancel out a huge density at the proper stage, and then to help
our universe in exiting from inflation. This is because the expansion
of a universe cannot dilute or deplete the vacuum energy density.
The solution raises new quandaries: What is the mechanism? Will any
relic of the vacuum energy density survive under the effect of this
mechanism?

From the statement above, one can imagine that people are attracted
and puzzled in equal measure by questions about the existence of the
cosmological constant and vacuum energy densities. As evidenced by
\cite{V&CCP-01,V&CCP-02,V&CCP-03,V&CCP-04,V&CCP-04-1,V&CCP-05,V&CCP-06,V&CCP-07,V&CCP-08,V&CCP-09,V&CCP-10,V&CCP-11,V&CCP-12,V&CCP-13,V&CCP-14,V&CCP-15}
and similar thinking that we have already mentioned, much work has
been proposed and undertaken around these questions. Impressively,
\cite{V&CCP-01} and several physicists have noted that the cosmological
constant can safely and gracefully coexist with general relativity
due to the elegant mathematics and the requirements of particle physics
theory. A voice \cite{V&CCP-14} was sounded in the spirit of \cite{V&CCP-01}
recently: Is it possible to set the cosmological term as a fundamental
constant like the speed of light $c$ and the Planck constant $\hbar$?
Regrettably, things are not so simple since two coincidental problems
cannot be abandoned. First, the second Friedman equation with a ``fundamental
Einstein's cosmological term $\lambda$'' in flat spacetime can be
written as 
\begin{equation}
H^{2}=\frac{\kappa c^{2}}{3}\left(\left\langle \varepsilon_{\mathrm{ord}}\right\rangle +\left\langle \varepsilon_{\mathrm{vac}}\right\rangle \right)+\frac{\lambda c^{2}}{3}\label{eq:1.5}
\end{equation}
(where $H=\nicefrac{\dot{\mathbb{R}}}{\mathbb{R}}$ is the Hubble
rate, and the total energy density is $\left\langle \varepsilon_{\mathrm{ord}}\right\rangle +\left\langle \varepsilon_{\mathrm{vac}}\right\rangle $,
which can be separated into two parts: the ordinary and the vacuum).
This asserts that the \textquotedblleft{}effective cosmological constant
(ECC) $\Lambda$\textquotedblright{} in density formation would be
\begin{equation}
\varepsilon_{\Lambda}=\left\langle \varepsilon_{\mathrm{vac}}\right\rangle +\frac{\lambda}{\kappa}.\label{eq:1.6}
\end{equation}
Indeed, a negative $\lambda$ could cancel the vacuum density out.
Nonetheless, the coincidence is too great for it to be identical to
the needed value when \eqref{eq:1.3} is compared to \eqref{eq:1.2}
and \eqref{eq:1.4}, as each of these is dependent on the chosen
inflationary theory. There is another coincidence too: that $\lambda$
will appear at a specific cosmic time, thereby fulfilling the condition
of inflationary theory that is employed to help our universe in ending
inflation (such as $t_{\mathrm{end}}\lesssim\unit[10^{-36}]{s}$ for
the inflation that begins when the universe's size is close to the
Planck point). The second problem is particularly curious: the \textquotedblleft{}fundamental
constant $\lambda$'' is most strange because it did not originally
exist and could not have appeared too early or too late, as otherwise
our universe would have turned out totally differently to the one
we see today.

Surely, if another mechanism without these coincidental problems could
be found, keeping the fundamental constant $\Lambda$ as (or close
to) \eqref{eq:1.3} would be perfectly acceptable.

Luckily, the Klein-Gordon equation edifies with the assertion that
scalar field dynamics could perform a process for decaying scalar
potential energy by self-interaction. \cite{V&CCP-02} and others
employed this concept and introduced the scalar field $\phi$, named
inflaton, as the quantum matter in the epoch before the phase transition
of Grand Unification Theory (GUT). Followingly, the Friedman equations
with only $\phi$ in flat spacetime are 
\begin{equation}
\frac{\ddot{\mathbb{R}}}{\mathbb{R}}=-\frac{\kappa}{3}\left(\dot{\phi}^{2}-c^{2}V\left(\phi\right)\right),\label{eq:1.7}
\end{equation}
\begin{equation}
\left(\frac{\dot{\mathbb{R}}}{\mathbb{R}}\right)^{2}=\frac{\kappa}{3}\left(\frac{1}{2}\dot{\phi}^{2}+c^{2}V\left(\phi\right)\right),\label{eq:1.8}
\end{equation}
where $\phi=\phi\left(t\right)$ is only dependent on time and $V\left(\phi\right)$
is the inflaton potential with a minimum value of zero. Clearly, $t_{i}$
sets the start time for inflation; $V\left(\phi\left(t_{i}\right)\right)$
is the vacuum energy density that triggers the inflation of the universe.
On the other hand, $V\left(\phi\left(t_{r}\right)\right)$ can be
treated as a cosmological constant when $\phi$ comes to rest at $t\geqslant t_{r}$.
To derive \eqref{eq:1.8} with respect to $t$, the field equation
in dynamic spacetime can be found as 
\begin{equation}
-3H\dot{\phi}^{2}=\frac{d}{dt}\left(\frac{1}{2}\dot{\phi}^{2}+c^{2}V\left(\phi\right)\right).\label{eq:1.9}
\end{equation}
Equation \eqref{eq:1.9} tells us that the damping term $3H\dot{\phi}^{2}$
will consume energy from $V\left(\phi\left(t_{i}\right)\right)$ when
a universe is expanding. Evidently, $V\left(\phi\left(t_{r}\right)\right)$
is zero since the total energy of $\dot{\phi}^{2}$ and $V\left(\phi\right)$
will be used up in the end. In this scenario, it looks as if setting
a nonzero minimum of $V\left(\phi\right)$, or allowing the insertion
of \textquotedblleft{}$\Lambda$\textquotedblright{} into \eqref{eq:1.7}
and \eqref{eq:1.8} are the only methods for obtaining the cosmological
constant.

\begin{figure}
\begin{centering}
\includegraphics[width=7cm]{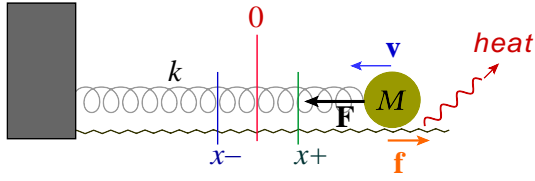} 
\par\end{centering}

\caption{\label{fig:1} A spring oscillating system is placed on a rough table: $\mathrm{k}$
is the spring constant; $\boldsymbol{\mathbf{F}}$ is the restoring
force of the spring; $\boldsymbol{\mathbf{f}}$ is the frictional
force between the oscillator and the table-surface; the mass of the
oscillator is $M$; ``$0$'' is the position of the spring's original
length; and $\left[x_{-},x_{+}\right]$ is the stagnant zone of the
oscillator.}
\end{figure}

Well, if one still believes that the cosmological constant is a \textquotedblleft{}product\textquotedblright{}
of the evolution of the universe, what can one do to search for this
mechanism? Actually, the scenario presented in \eqref{eq:1.7}
and \eqref{eq:1.8} is too simple because it only includes quantum
matter. Such limitations lead to another question: Is it possible
that the process of creating matter in the early universe could hint
towards the building mechanism for the cosmological constant? Following
the above conjecture and the fact that \eqref{eq:1.9} is an oscillating
equation, there is a simple but useful example that arises in daily
life: a spring oscillating system on the rough surface of a table,
as in Figure \ref{fig:1} . 

\begin{figure}
\begin{centering}
\includegraphics[width=9cm]{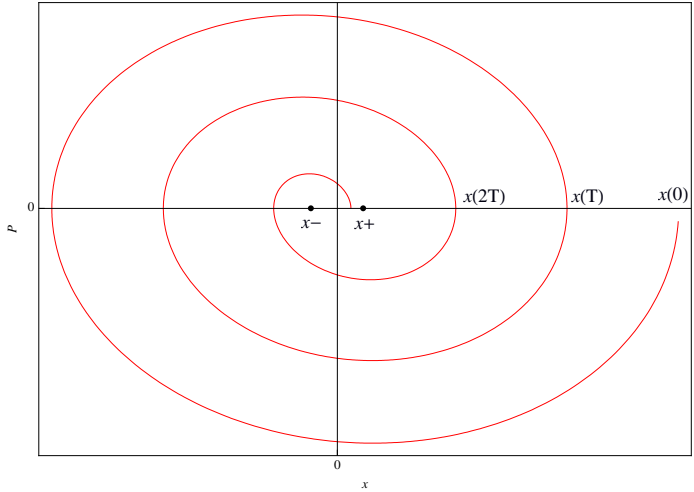}
\par\end{centering}

\caption{\label{fig:2} This is the phase diagram cited in the example. According to \eqref{eq:1.12},
$x_{+}$ is the center of the elliptical path of the bottom part;
$x_{-}$ is the other center of the upper part, obtained from \eqref{eq:1.13}.
It is clear that the oscillator will stop forever at the position
inside the stagnant zone when its amplitude is not bigger than $\left\Vert x_{\pm}-0\right\Vert $.}
\end{figure}

If the amount of kinetic frictional force is set equal to the maximum
static frictional force, the equations of motion will be
\begin{equation}
M\ddot{x}+\mathrm{k}x=+f,\quad\mathrm{for}\:\dot{x}<0,\label{eq:1.10}
\end{equation}
\begin{equation}
M\ddot{x}+\mathrm{k}x=-f,\quad\mathrm{for}\:\dot{x}>0.\label{eq:1.11}
\end{equation}
$t$ denotes the time of lab frame and $f$ is the amount of frictional
force between the table-surface and the oscillator, $M$. \eqref{eq:1.10}
can now be rewritten as a phase equation with the moment $\boldsymbol{\mathbf{P}}$
and position $\boldsymbol{\mathbf{x}}$: 
\begin{equation}
\frac{P^{2}\left(t\right)}{M\mathrm{k}\left(x\left(n\mathrm{T}\right)-\frac{f}{\mathrm{k}}\right)^{2}}+\frac{\left(x\left(t\right)-\frac{f}{\mathrm{k}}\right)^{2}}{\left(x\left(n\mathrm{T}\right)-\frac{f}{\mathrm{k}}\right)^{2}}=1,\label{eq:1.12}
\end{equation}
$\mathrm{T}=2\pi\sqrt{\frac{M}{\mathrm{k}}}$ is the period of oscillation.
This equation only applies to the interval of time $n\mathrm{T}\leq t\leq\frac{2n+1}{2}\mathrm{T}$,
where $n=0,1,2,\cdots$. In addition, \eqref{eq:1.11} becomes
\begin{equation}
\frac{P^{2}\left(t\right)}{M\mathrm{k}\left(x\left(\frac{2n+1}{2}\mathrm{T}\right)+\frac{f}{\mathrm{k}}\right)^{2}}+\frac{\left(x\left(t\right)+\frac{f}{\mathrm{k}}\right)^{2}}{\left(x\left(\frac{2n+1}{2}\mathrm{T}\right)+\frac{f}{\mathrm{k}}\right)^{2}}=1,\label{eq:1.13}
\end{equation}
which is available during $\frac{2n+1}{2}\mathrm{T}\leq t\leq\left(n+1\right)\mathrm{T}$,
where $n=0,1,2,\cdots$. Figure \ref{fig:2}  
 and the stagnant zone, $\left[x_{+}=\frac{+f}{\mathrm{k}},x_{-}=\frac{-f}{\mathrm{k}}\right]$,
can be drawn through \eqref{eq:1.12} and \eqref{eq:1.13}.

We will see that the nonzero oscillator's potential might survive
if the oscillator itself rests at a position that is not the origin
inside the stagnant zone.

We can now assume certain kinds of \textquotedblleft{}effective friction\textquotedblright{}
and introduce them to \eqref{eq:1.9}. Moreover, the part of radiation
that should be inserted into \eqref{eq:1.7} and \eqref{eq:1.8}
comes as the result of the work done by ``effective kinetic frictional
force''. This is similar to the above example, and it will hopefully
help us to obtain the \textquotedblleft{}stagnant zone\textquotedblright{}
of $\phi$ and then to find the final range of the remaining potential
$V\left(\phi\left(t_{r}\right)\right)$.

Inspired by the example of the spring oscillating system, we will
now develop our idea in order to build the cosmological constant.
The following paper is organized in this way: Firstly, we will illustrate
the energy relation between inflaton $\phi$ and radiation (relativistic
particles) in a homogeneous and isotropic universe. Next, according
to the settings of Section \ref{sec:2}, we will introduce two fundamental postulates
to build the theory of thermo-inflation. In Section \ref{sec:4}, from the order
of radiation creation discussed in Section \ref{sec:2}, we deduce that \textquotedblleft{}effective
frictional force\textquotedblright{} can exist in inflaton dynamics,
and use this result to help discover the relic of inflaton potential.
Then, by considering the conditions of \textquotedblleft{}effective
friction\textquotedblright{}, two special types of universe will be
shown: a Type I universe will finally enter an expanding course with
uniformly rolling $\dot{\phi}\left(t_{*}\right)$ after $t_{*}$,
and particles will be created continuously during this course; in
a Type II universe, a relic of inflaton potential, $V\left(\phi\left(t_{r}\right)\right)$,
will survive to become the ECC $\Lambda$ when $\phi$ is at rest
after $t_{r}$. Conclusions will be provided in Section \ref{sec:5}.

It should be noted that, in the following sections, we use the Planck
units: $c=G=k_{\mathrm{B}}=\hbar=1$. Unless specifically mentioned,
$t$, $\tau$ and $\eta$ are employed as the cosmic time, and $t_{i}$
denotes the point at which inflation was beginning and $\phi$ comes
to rest at $t\geqslant t_{r}$.

\section{The energy relation between inflaton and radiation} \label{sec:2}

\subsection{Models} \label{sec:2.1}

First of all, the energy-momentum tensors of the material in the universe
must be recorded. Since the epoch that interests us is earlier than
GUT phase transition, most of the material at this time can be considered
as a quantum matter, inflaton $\phi\left(x^{\mu}\right)$, which is
a kind of scalar field. In our scenario, $\phi$ is real and its energy-momentum
tensor is 
\begin{equation}
T_{\mu\nu}^{\left(\phi\right)}=\partial_{\mu}\phi\partial_{\nu}\phi-g_{\mu\nu}\left[\frac{1}{2}g^{\alpha\beta}\partial_{\alpha}\phi\partial_{\beta}\phi-V\left(\phi\right)\right],\label{eq:2.1.1}
\end{equation}
where $V\left(\phi\right)$ is the inflaton potential. In addition,
assuming that radiation can simultaneously exist, its energy-momentum
tensor should be 
\begin{equation}
T_{\mu\nu}^{\left(\mathrm{r}\right)}=\frac{\varepsilon_{\mathrm{r}}\left(t\right)}{3}\left(4u_{\mu}u_{\nu}-g_{\mu\nu}\right)\label{eq:2.1.2}
\end{equation}
owing to the pressure of radiation $p_{\mathrm{r}}=\nicefrac{\varepsilon_{\mathrm{r}}}{3}$
(where $u_{\mu}$ is the 4-velocity and $\varepsilon_{\mathrm{r}}$
is the energy density of radiation).

Once this is complete, the distribution of inflaton and radiation
in spacetime, and the spacetime geometry of the very early universe
are also required. Considering our goal is to search for the mechanism
that builds the cosmological constant, the equation of general relativity
without the cosmological term should be used, as
\begin{equation}
R_{\mu\nu}-\frac{1}{2}Rg_{\mu\nu}=-16\pi\underset{j}{\sum}T_{\mu\nu}^{\left(j\right)},\label{eq:2.1.3}
\end{equation}
where $R$ is the Ricci scalar, and $\underset{j}{\sum}T_{\mu\nu}^{\left(j\right)}$
is the total energy-momentum tensor. According to observations, our
universe is homogeneous and isotropic on large scales, meaning that
the left hand side of the equal sign in \eqref{eq:2.1.3} should
be very close to the off-diagonal. For the sake of simplicity, each
individual component of $T_{\mu\nu}^{\left(j\right)}$ is undoubtedly
off-diagonal as well. Therefore, the distribution of radiation in
flat spacetime is 
\begin{equation}
\left[u_{\mu}\right]=\left(1,0,0,0\right),\label{eq:2.1.4}
\end{equation}
in which the mean velocity of radiation in an arbitrary spatial direction
is $u_{k}=0$ to satisfy the requirement of there being no net current
of matter in the universe. Furthermore, setting $\phi\left(x^{\mu}\right)=\phi\left(t\right)$
to guarantee that $T_{\mu\nu}^{\left(\phi\right)}$ will not violate
actual observations of our universe, the components of $T_{\mu\nu}^{\left(\phi\right)}$
are 
\begin{equation}
\begin{array}{l}
T_{00}^{\left(\phi\right)}=\varepsilon_{\phi}\left(t\right)=\frac{1}{2}\dot{\phi}^{2}+V,\\
T_{kk}^{\left(\phi\right)}=p_{\phi}\left(t\right)=\frac{1}{2}\dot{\phi}^{2}-V,\\
T_{kl}^{\left(\phi\right)}=0,\;\left(k\neq l\right).
\end{array}\label{eq:2.1.5}
\end{equation}
Finally, we assume that the initial universe is flat, as consistent
with our previous discussion. It means that the FRW line element should
be 
\begin{equation}
ds^{2}=dt^{2}-\mathbb{R}^{2}\left(t\right)\left(dx^{2}+dy^{2}+dz^{2}\right),\label{eq:2.1.6}
\end{equation}
as the background of the universe's spacetime.

\subsection{Energy conservation and interaction} \label{sec:2.2}

Noether taught us that an isolated physical system should obey the
requirement of its total action being invariant under an infinitesimal
coordinate transformation. This leads to a conservation law of
\begin{equation}
D_{\mu}\left(\underset{j}{\sum}T_{\left(j\right)}^{\mu\nu}\right)=0\label{eq:2.2.1}
\end{equation}
in curved spacetime, where $D_{\mu}$ is the covariant derivative
and $\sum_{j}$ means a system with several matter fields. By this
reasoning, if we assume that our universe is unique or adiabatic,
as per the discussion in the previous subsection, $D_{\mu}\left(T_{\left(\mathrm{r}\right)}^{\mu\nu}+T_{\left(\phi\right)}^{\mu\nu}\right)=0$
must be satisfied. For $\nu=0$, the energy conservation is obeyed
by 
\begin{equation}
\dot{\varepsilon}_{\mathrm{r}}+4H\varepsilon_{\mathrm{r}}=-\left(\dot{\varepsilon}_{\phi}+3H\dot{\phi}^{2}\right),\label{eq:2.2.2}
\end{equation}
which clearly shows the relationship of energy transference between
$\phi$ and radiation. Thus we further consider the situation of interaction,
and \eqref{eq:2.2.2} can be separated into two situations:
\begin{description}
\item [{Situation 1:}] There is no interaction between $\phi$
and radiation. This is popular for discussion purposes. In this situation,
\eqref{eq:2.2.2} obeys 
\begin{equation}
\dot{\varepsilon}_{\mathrm{r}}+4H\varepsilon_{\mathrm{r}}=0,\quad\dot{\varepsilon}_{\phi}+3H\dot{\phi}^{2}=0.\label{eq:2.2.3}
\end{equation}
Obviously, the components of both radiation and $\phi$ are closed,
since energy cannot transfer between the two.
\item [{Situation 2:}] Some interaction exists between $\phi$
and radiation. A similar property is first considered for the case
of quintessence and matter by Zimdahl et al. \cite{mod(IQ)-01}. In
this situation, the interaction term $Q\left(t\right)$ can be introduced
into \eqref{eq:2.2.2} as 
\begin{equation}
\dot{\varepsilon}_{\mathrm{r}}+4H\varepsilon_{\mathrm{r}}=Q\left(t\right),\quad\dot{\varepsilon}_{\phi}+3H\dot{\phi}^{2}=-Q\left(t\right).\label{eq:2.2.4}
\end{equation}
It follows that $\phi$ and radiation are open to each other.
\end{description}

\section{Theory of thermo-inflation} \label{sec:3}

\subsection{Postulates} \label{sec:3.1}

The properties for creating radiation before and after GUT phase transition
must be outlined. We suggest following postulates for our scenario:
\begin{description}
\item [{Postulate A:}] There exists some interaction between $\phi$
and radiation%
\footnote{Similar consideration applied to the models of quintessence and phantom
were proposed by \cite{mod(IQ)-01,mod(IQ)-02,mod(IQ)-03,mod(IQ)-04}.%
}. 
\item [{Postulate B:}] Radiation can be created continuously during
and after the epoch of inflation. \cite{TI-01,TI-02}
\end{description}
Alternatively, when considering the creation of radiation in an adiabatic
universe, Prigogine et al. \cite{TC} suggest the thermal condition
of an open system of radiation 
\begin{equation}
d\left(\varepsilon_{\mathrm{r}}v\right)+p_{\mathrm{r}}dv-\frac{h_{\mathrm{r}}}{n_{\mathrm{r}}}d\left(n_{\mathrm{r}}v\right)=0,\label{eq:3.1.1}
\end{equation}
which obeys the first law of thermodynamics ($h_{\mathrm{r}}=\varepsilon_{\mathrm{r}}+p_{\mathrm{r}}=\frac{4}{3}\varepsilon_{\mathrm{r}}$;
$v\propto\mathbb{R}^{3}$ is the comoving volume of the universe;
$n_{\mathrm{r}}\left(t\right)=\nicefrac{N_{\mathrm{r}}\left(t\right)}{v\left(t\right)}$
is the number density of relativistic particles; and $N_{\mathrm{r}}\left(t\right)$
is the number of particles in the whole universe). \eqref{eq:3.1.1}
has another easily calculable form as 
\begin{equation}
\dot{\varepsilon}_{\mathrm{r}}+4H\varepsilon_{\mathrm{r}}=\frac{4}{3}\Gamma\varepsilon_{\mathrm{r}},\label{eq:3.1.2}
\end{equation}
where $\Gamma\left(t\right)\equiv\nicefrac{\dot{N}_{\mathrm{r}}}{N_{\mathrm{r}}}$
is the particle creation rate. The solution of radiation energy density
$\varepsilon_{\mathrm{r}}$ is 
\begin{equation}
\varepsilon_{\mathrm{r}}\left(t\right)=\varepsilon_{\mathrm{r}}\left(t_{a}\right)\exp\left(\frac{4}{3}\int_{t_{a}}^{t}\left(\Gamma-3H\right)d\tau\right),\label{eq:3.1.3}
\end{equation}
where $t_{a}$ is an arbitrary cosmic time for the commencement of
observation%
\footnote{Alternatively, the particle-number at time $t$ can be solved from \eqref{eq:3.1.2}, as 
\begin{equation}
N_{\mathrm{r}}\left(t\right)=\chi\left(\varepsilon_{\mathrm{r}}\left(t\right)\mathbb{R}^{4}\left(t\right)\right)^{\nicefrac{3}{4}}.\label{eq:3.1.4}
\end{equation}
Given that the particles of radiation could be divided into bosons
and fermions, the constant $\chi$ can be found as 
\begin{equation}
\chi=\frac{n_{\mathrm{b}}\left(t\right)+n_{\mathrm{f}}\left(t\right)}{\left[\varepsilon_{\mathrm{b}}\left(t\right)+\varepsilon_{\mathrm{f}}\left(t\right)\right]^{\nicefrac{3}{4}}}=\frac{\frac{\zeta\left(3\right)}{\pi^{2}}\left(\underset{\mathrm{b}}{\sum}g_{\mathrm{b}}T_{\mathrm{b}}^{3}\left(t\right)+\frac{3}{4}\underset{\mathrm{f}}{\sum}g_{\mathrm{f}}T_{\mathrm{f}}^{3}\left(t\right)\right)}{\left[\frac{\pi^{2}}{30}\left(\underset{\mathrm{b}}{\sum}g_{\mathrm{b}}T_{\mathrm{b}}^{4}\left(t\right)+\frac{7}{8}\underset{\mathrm{f}}{\sum}g_{\mathrm{f}}T_{\mathrm{f}}^{4}\left(t\right)\right)\right]^{\nicefrac{3}{4}}}\label{eq:3.1.5}
\end{equation}
according to the Bose-Einstein and Fermi-Dirac distributions. Here
$\zeta\left(3\right)=1.20206\ldots$ is the Riemann zeta function
of $3$; $g_{\mathrm{b}}$ and $g_{\mathrm{f}}$ are degrees of freedom
for bosons and fermions; $T_{\mathrm{b}}\left(t\right)$ and $T_{\mathrm{f}}\left(t\right)$
are the temperatures of bosons and fermions. In addition, $T_{\mathrm{b,\, f}}\left(t\right)\gg m_{\mathrm{b,\, f}}$
and $T_{\mathrm{b,\, f}}\left(t\right)\gg\mu_{\mathrm{b,\, f}}$ during
the radiation-dominated era (where $m$ is the mass of particle, and
$\mu$ is the chemical potential). If an epoch of thermal equilibrium
is discovered somewhere in the history of the radiation-dominated
era, the temperature terms of \eqref{eq:3.1.5} can be eliminated.%
}. Besides, comparing \eqref{eq:3.1.2} with \eqref{eq:2.2.2},
$\Gamma$ could be described as 
\begin{equation}
\Gamma=-\frac{\dot{\varepsilon}_{\phi}+3H\dot{\phi}^{2}}{\frac{4}{3}\varepsilon_{\mathrm{r}}}.\label{eq:3.1.6}
\end{equation}
Because of Postulate B, the value of $\Gamma$ should \textbf{NOT}
be less than zero. It leads to the fact that $\dot{\varepsilon}_{\phi}+3H\dot{\phi}^{2}\leq0$
(the interaction term $Q\left(t\right)=\frac{4}{3}\Gamma\varepsilon_{\mathrm{r}}\geq0$)
happens in an expanding universe. Therefore, in our scenario, the energy of $\phi$ 
will decrease with time and flow into radiation creation.
In other words, the discovery of \eqref{eq:3.1.2} gives us an
important and specific message: if we believe that our universe was
created from a vacuum, the energy relationship between quantum matter
and real matter should satisfy Situation 2.

Introducing \eqref{eq:2.1.5} into \eqref{eq:3.1.6}, we easily
find that radiation is created by the moving $\phi$. Moreover, the
following equation is relevant, 
\begin{equation}
\underset{\dot{\phi}\rightarrow0}{\lim}\Gamma\left(t\right)=-\underset{\dot{\phi}\rightarrow0}{\lim}\frac{\left(\ddot{\phi}+V^{'}\left(\phi\right)+3H\dot{\phi}\right)\dot{\phi}}{\frac{4}{3}\varepsilon_{\mathrm{r}}}=0,\label{eq:3.1.7}
\end{equation}
where $V^{'}\left(\phi\right)=\frac{dV}{d\phi}$. Consequently there
is no radiation created when the \textquotedblleft{}instantaneous
speed\textquotedblright{} of $\phi$ is zero. Of course, radiation
will no longer be created when $t\geq t_{r}$.

\subsection{Field equations and solutions} \label{sec:3.2}

Relying on \eqref{eq:2.1.3}, the Friedman equations with field
$\phi$ and radiation can be written as 
\begin{equation}
\frac{\ddot{\mathbb{R}}}{\mathbb{R}}=-\frac{8\pi}{3}\left(\dot{\phi}^{2}-V\left(\phi\right)+\varepsilon_{\mathrm{r}}\right),\label{eq:3.2.1}
\end{equation}
\begin{equation}
\left(\frac{\dot{\mathbb{R}}}{\mathbb{R}}\right)^{2}=\frac{8\pi}{3}\left(\frac{1}{2}\dot{\phi}^{2}+V\left(\phi\right)+\varepsilon_{\mathrm{r}}\right).\label{eq:3.2.2}
\end{equation}
The employment of the relationship $p_{\mathrm{r}}=\nicefrac{\varepsilon_{\mathrm{r}}}{3}$
is again worthy of attention. To derive both sides of the equal sign
in \eqref{eq:3.2.2} with respect to $t$, and to combine our
result with \eqref{eq:3.2.1}, we have 
\begin{equation}
-H\left(3\dot{\phi}^{2}+4\varepsilon_{\mathrm{r}}\right)=\frac{d}{dt}\left(\frac{1}{2}\dot{\phi}^{2}+V+\varepsilon_{\mathrm{r}}\right)=\frac{3}{4\pi}H\dot{H}.\label{eq:3.2.3}
\end{equation}
It gives the solution of the Hubble rate 
\begin{equation}
H\left(t\right)=H\left(t_{i}\right)-\frac{4\pi}{3}\int_{t_{i}}^{t}\left(3\dot{\phi}^{2}+4\varepsilon_{\mathrm{r}}\right)d\tau,\label{eq:3.2.4}
\end{equation}
and therefore the spatial scale factor is 
\begin{equation}
\mathbb{R}\left(t\right)=\mathbb{R}\left(t_{i}\right)\exp\left[\int_{t_{i}}^{t}\left(H\left(t_{i}\right)-\frac{4\pi}{3}\int_{t_{i}}^{\eta}\left(3\dot{\phi}^{2}+4\varepsilon_{\mathrm{r}}\right)d\tau\right)d\eta\right].\label{eq:3.2.5}
\end{equation}
\eqref{eq:3.2.3} also provides the evolution of radiation density:
\begin{equation}
\varepsilon_{\mathrm{r}}\left(t\right)=-\frac{3}{16\pi}\dot{H}-\frac{3}{4}\dot{\phi}^{2}.\label{eq:3.2.6}
\end{equation}
This is the other form of \eqref{eq:3.1.3}. Next, comparing \eqref{eq:3.2.1}
with the second derivative of \eqref{eq:3.2.5} (with respect
to $t$), a solution for the inflaton potential becomes clear: 
\begin{equation}
V\left(t\right)=\frac{3}{8\pi}\left[H\left(t_{i}\right)-\frac{4\pi}{3}\int_{t_{i}}^{t}\left(3\dot{\phi}^{2}+4\varepsilon_{\mathrm{r}}\right)d\tau\right]^{2}-\frac{1}{2}\dot{\phi}^{2}\left(t\right)-\varepsilon_{\mathrm{r}}\left(t\right),\label{eq:3.2.7}
\end{equation}
or 
\begin{equation}
V\left(t\right)=\frac{1}{4}\dot{\phi}^{2}+\frac{R}{32\pi},\label{eq:3.2.8}
\end{equation}
where $R=6\dot{H}+12H^{2}$ is the Ricci curvature. Attention should
be drawn to the fact that $V\left(t\right)$ is the expansion in $t$
of $V\left(\phi\left(t\right)\right)$; it is not $V\left(\phi,t\right)$.

\section{``Effective friction'' and the relic of inflaton potential} \label{sec:4}

\subsection{Definition of the effective friction} \label{sec:4.1}

In this section, our mission is to demonstrate that an effect similar
to friction could exist in inflaton dynamics. Suppose that the inflaton
is affected by the effect, its equation of motion will be
\begin{equation}
\ddot{\phi}+3H\dot{\phi}+V^{'}\left(\phi\right)=\pm f_{\phi k},\label{eq:4.1.0}
\end{equation}
where the source term $f_{\phi k}$ is the \textquotedblleft{}effect/force\textquotedblright{}
that we assume and wish to find out. In addition, the direction of
the force is defined by ``$+$'' or ``$-$''. The ``$+$'' direction
follows the situation of $\dot{\phi}<0$; \textquotedblleft{}$-$\textquotedblright{}
will be adopted if $\dot{\phi}>0$.

Next, in order to obtain the equation as \eqref{eq:4.1.0}, we
introduce \eqref{eq:3.1.2} into the calculating result of \eqref{eq:3.2.3}.
We obtain 
\begin{equation}
\dot{\phi}\ddot{\phi}+3H\dot{\phi}^{2}+V^{'}\left(\phi\right)\dot{\phi}=-\frac{4}{3}\Gamma\varepsilon_{\mathrm{r}}.\label{eq:4.1.1}
\end{equation}
According to the discussion of \eqref{eq:3.1.6}, the negative
sign of \eqref{eq:4.1.1} shows that energy is transferred from
$\phi$ to radiation. Taking the macroscopic viewpoint, and comparing
\eqref{eq:4.1.1} with \eqref{eq:1.10} and \eqref{eq:1.11},
$-\frac{4}{3}\Gamma\varepsilon_{\mathrm{r}}$ can be regarded as the
power resulting from some ``kinetic frictional force'', just like
the example in Section \ref{sec:1}. Rewriting \eqref{eq:4.1.1} as 
\begin{equation}
\ddot{\phi}+3H\dot{\phi}+V^{'}\left(\phi\right)=-\frac{4\Gamma\varepsilon_{\mathrm{r}}}{3\dot{\phi}},\label{eq:4.1.2}
\end{equation}
we have the following conclusion after comparing \eqref{eq:4.1.2}
with \eqref{eq:4.1.0}:
\begin{equation}
\mathbf{f}_{\phi k}\equiv-\frac{4\Gamma\varepsilon_{\mathrm{r}}}{3\dot{\phi}}.\label{eq:4.1.2.2}
\end{equation}
Here $\mathbf{f}_{\phi k}$ is the vector-form of $f_{\phi k}$, and
its direction is indeed opposite to the direction of $\dot{\phi}$
due to the conclusion of the positive interaction term $Q\left(t\right)$
from \eqref{eq:2.2.4} and \eqref{eq:3.1.6}. Therefore, $\mathbf{f}_{\phi k}$
can be defined as the \textquotedblleft{}Effective Kinetic Frictional
Force\textquotedblright{} (EKFF) of the oscillating system of $\phi$
($\phi$-system for short). Furthermore, \eqref{eq:4.1.2} provides
the additional information of the \textquotedblleft{}Effective Static
Frictional Force\textquotedblright{} (ESFF), as

\begin{equation}
\left\Vert \boldsymbol{\mathbf{f}}_{\phi s}\right\Vert =\underset{\dot{\phi}\rightarrow0}{\lim}\left\Vert -\frac{4\Gamma\varepsilon_{\mathrm{r}}}{3\dot{\phi}}\right\Vert .\label{eq:4.1.3}
\end{equation}
At first glance, $\left\Vert \boldsymbol{\mathbf{f}}_{\phi s}\right\Vert $
seems to cause confusion because it looks like divergent. In actual
fact, this is not a matter for worry because \textbf{a static frictional
force does zero work}. This means that a finite $\left\Vert \boldsymbol{\mathbf{f}}_{\phi s}\right\Vert $
can exist in the $\phi$-system as per the discussion of \eqref{eq:3.1.7}.

\subsection{Two special types of universe} \label{sec:4.2}

To apply the conclusion of \textquotedblleft{}effective friction\textquotedblright{},
the universe can be roughly sorted by two special conditions of \eqref{eq:4.1.2}
and \eqref{eq:4.1.3}. These conditions and the corresponding
universe types are:
\begin{enumerate}
\item Type I universe --- expanding with an uniform $\dot{\phi}\left(t_{*}\right)$:
For a variable EKFF, it and the damping term $-3H\left(t\right)\dot{\phi}\left(t_{*}\right)$
will cancel out the restoring force $-V^{'}\left(\phi\left(t\right)\right)$
after the special time $t_{*}$, causing $\dot{\phi}$ to become the
uniform velocity $\dot{\phi}\left(t_{*}\right)$. In consequence,
the universe will enter an expanding mode as 
\begin{equation}
\mathbb{R}\left(t\right)=\mathbb{R}\left(t_{*}\right)\exp\left[-\int_{t_{*}}^{t}\left(\frac{4\Gamma\left(\tau\right)\varepsilon_{\mathrm{r}}\left(\tau\right)}{9\dot{\phi}^{2}\left(t_{*}\right)}+\frac{V^{'}\left(\phi\left(\tau\right)\right)}{3\dot{\phi}\left(t_{*}\right)}\right)d\tau\right].\label{eq:4.2.1-1}
\end{equation}
Another important property of this type can also be discovered: particles
would have had a period of creation after $t_{*}$. The range of the
particle creation rate should be 
\begin{equation}
0<\Gamma\left(t\right)<-\frac{3V^{'}\left(\phi\left(t\right)\right)\dot{\phi}\left(t_{*}\right)}{4\varepsilon_{\mathrm{r}}\left(t\right)},\label{eq:4.2.1-2}
\end{equation}
where $t\geq t_{*}$. However, this will be invalid after the universe
enters the next stage (such as $\dot{\phi}$ will not be uniform).
\begin{description}
\item [{Example:}] We are drawn to Hoyle\textquoteright{}s 1948 model \cite{Hoyle}
since a Type I universe can create particles continuously even when
its age is much older than the epoch of inflation. Now, let us consider
a toy condition in which \emph{the matter of a special Type I universe
is always radiation}. The Hubble rate of such a universe can actually
be found from \eqref{eq:3.2.6}, as
\begin{equation}
H\left(t\right)=H\left(t_{\ddagger}\right)-\frac{4\pi}{3}\left(3\dot{\phi}^{2}\left(t_{*}\right)+4\varepsilon_{\mathrm{r}}\left(t_{\ddagger}\right)\right)\left(t-t_{\ddagger}\right),\label{eq:4.2.1-3}
\end{equation}
which satisfies the requirement that the matter density of Hoyle\textquoteright{}s
universe will become static after $t_{\ddagger}$ (the time-relationship
is $t\geq t_{\ddagger}\geq t_{*}$). In this case, \eqref{eq:4.1.2}
and \eqref{eq:4.2.1-3} provide an interesting and important result
of
\begin{eqnarray}
3\dot{\phi}\left(t_{*}\right)V^{'}\left(\phi\left(t\right)\right)+4\varepsilon_{\mathrm{r}}\left(t_{\ddagger}\right)\Gamma\left(t\right)=\nonumber \\
-9\dot{\phi}^{2}\left(t_{*}\right)H\left(t_{\ddagger}\right)+12\pi\dot{\phi}^{2}\left(t_{*}\right)\left(3\dot{\phi}^{2}\left(t_{*}\right)+4\varepsilon_{\mathrm{r}}\left(t_{\ddagger}\right)\right)\left(t-t_{\ddagger}\right)\label{eq:4.2.1-4}
\end{eqnarray}
since $\ddot{\phi}\left(t\geq t_{*}\right)=0$. In addition, to introduce
\eqref{eq:4.2.1-3} into \eqref{eq:3.2.2} with conditions
of $\dot{\phi}\left(t\right)=\dot{\phi}\left(t_{*}\right)$, $\varepsilon_{\mathrm{r}}\left(t\right)=\varepsilon_{\mathrm{r}}\left(t_{\ddagger}\right)$
and $V\left(\phi\right)=V\left(\phi\left(t\right)\right)$, the value
of inflaton potential at $t$ can be obtained as
\begin{eqnarray}
V\left(\phi\left(t\right)\right) & = & \frac{3}{8\pi}\left[H\left(t_{\ddagger}\right)-\frac{4\pi}{3}\left(3\dot{\phi}^{2}\left(t_{*}\right)+4\varepsilon_{\mathrm{r}}\left(t_{\ddagger}\right)\right)\left(t-t_{\ddagger}\right)\right]^{2}\nonumber \\
 &  & -\frac{1}{2}\dot{\phi}^{2}\left(t_{*}\right)-\varepsilon_{\mathrm{r}}\left(t_{\ddagger}\right).\label{eq:4.2.1-5}
\end{eqnarray}
It looks very much like $V\left(\phi\left(t\right)\right)=\frac{1}{2}m_{\phi}^{2}\phi^{2}\left(t\right)+V_{0}$.
Comparing coefficients of $\left(t-t_{\ddagger}\right)^{n}$ after
we substitute $\phi\left(t\right)$ of $V\left(\phi\left(t\right)\right)=\frac{1}{2}m_{\phi}^{2}\phi^{2}\left(t\right)+V_{0}$
with $\phi\left(t_{\ddagger}\right)+\dot{\phi}\left(t_{*}\right)\left(t-t_{\ddagger}\right)$,
we obtain
\begin{equation}
m_{\phi}^{2}=\frac{4\pi}{3}\left(\frac{3\dot{\phi}^{2}\left(t_{*}\right)+4\varepsilon_{\mathrm{r}}\left(t_{\ddagger}\right)}{\dot{\phi}\left(t_{*}\right)}\right)^{2},\label{eq:4.2.1-6}
\end{equation}
\begin{equation}
\phi\left(t_{\ddagger}\right)=\frac{-3\dot{\phi}\left(t_{*}\right)H\left(t_{\ddagger}\right)}{4\pi\left(3\dot{\phi}^{2}\left(t_{*}\right)+4\varepsilon_{\mathrm{r}}\left(t_{\ddagger}\right)\right)},\label{eq:4.2.1-7}
\end{equation}
\begin{equation}
V_{0}=-\frac{1}{2}\dot{\phi}^{2}\left(t_{*}\right)-\varepsilon_{\mathrm{r}}\left(t_{\ddagger}\right).\label{eq:4.2.1-8}
\end{equation}
These results confirm our conjecture, but to our surprise, we discover
that the minimum potential $V_{0}$ of the Type I-Hoyle universe is
a negative and characteristic value as \eqref{eq:4.2.1-8}. Additionally,
the particle creation rate can be obtained by calculating \eqref{eq:4.2.1-4}
with \eqref{eq:4.2.1-6} and \eqref{eq:4.2.1-7}, as
\begin{eqnarray}
\Gamma\left(t\right) & = & 3H\left(t_{\ddagger}\right)-4\pi\left(3\dot{\phi}^{2}\left(t_{*}\right)+4\varepsilon_{\mathrm{r}}\left(t_{\ddagger}\right)\right)\left(t-t_{\ddagger}\right)\nonumber \\
 & = & 3H\left(t\right)\label{eq:4.2.1-9}
\end{eqnarray}
due to the fact that $V^{'}\left(\phi\left(t\right)\right)=m_{\phi}^{2}\phi\left(t\right)$.
Returning to the discussion of the Hubble rate, because of the $H\left(t\right)>0$
necessity, the lifetime of the Type I-Hoyle universe obeys
\begin{equation}
t_{\mathrm{LF}}<t_{\ddagger}+\frac{3H\left(t_{\ddagger}\right)}{4\pi\left(3\dot{\phi}^{2}\left(t_{*}\right)+4\varepsilon_{\mathrm{r}}\left(t_{\ddagger}\right)\right)}.\label{eq:4.2.1-10}
\end{equation}
On the other hand, if the universe is expanding with acceleration
during the Hoyle course, employing \eqref{eq:3.2.1} with conditions
of $\dot{\phi}\left(t\right)=\dot{\phi}\left(t_{*}\right)$, $\varepsilon_{\mathrm{r}}\left(t\right)=\varepsilon_{\mathrm{r}}\left(t_{\ddagger}\right)$
and $V\left(\phi\right)=V\left(\phi\left(t\right)\right)$, we obtain
\begin{equation}
t_{\mathrm{A}}<t_{\ddagger}+\frac{3H\left(t_{\ddagger}\right)}{4\pi\left(3\dot{\phi}^{2}\left(t_{*}\right)+4\varepsilon_{\mathrm{r}}\left(t_{\ddagger}\right)\right)}-\sqrt{\frac{3}{4\pi\left(3\dot{\phi}^{2}\left(t_{*}\right)+4\varepsilon_{\mathrm{r}}\left(t_{\ddagger}\right)\right)}}.\label{eq:4.2.1-11}
\end{equation}
Here, $t_{\mathrm{A}}$ is the age of accelerating expansion, and
the right hand part of ``$<$'' is the upper limit of $t_{\mathrm{A}}$.
In other words, a Type I universe will not have accelerating expansion
when it enters the Hoyle course, if the Hubble rate at $t_{\ddagger}$
satisfies
\begin{equation}
H\left(t_{\ddagger}\right)<\sqrt{\frac{4\pi}{3}\left(3\dot{\phi}^{2}\left(t_{*}\right)+4\varepsilon_{\mathrm{r}}\left(t_{\ddagger}\right)\right)}.\label{eq:4.2.1-12}
\end{equation}
\end{description}

\item Type II universe --- a relic, $V\left(\phi\left(t_{r}\right)\right)$,
will survive: For easy imaging, assume that the amount of EKFF and
the \textquotedblleft{}Maximum Effective Static Frictional Force\textquotedblright{}
(MESFF) are identical and invariable during each stage of the universe.
Consequently, according to \eqref{eq:4.1.2}, there are two positions
of $\phi$ where the net force is zero. They are 
\begin{equation}
V^{'}\left(\phi_{+}\right)=+f_{\phi}+3H\left\Vert \dot{\phi}_{+}\right\Vert ,\quad\mathrm{for}\;\dot{\phi_{+}}<0,\label{eq:4.2.2-2}
\end{equation}
\begin{equation}
V^{'}\left(\phi_{-}\right)=-f_{\phi}-3H\left\Vert \dot{\phi}_{-}\right\Vert ,\quad\mathrm{for}\;\dot{\phi}_{-}>0.\label{eq:4.2.2-3}
\end{equation}
Here $f_{\phi}$ denotes the amount of EKFF and MESFF. Dependent on
these results, the restoring force $V^{'}\left(\phi\left(t\right)\right)$
will be restricted between $-f_{\phi}$ and $+f_{\phi}$ when $t\geq t_{r}$.
By way of explanation, if the restoring force is no longer bigger
than the MESFF, $\phi$ will always stop at $\phi\left(t_{r}\right)$
inside a region named the \textquotedblleft{}stagnant zone\textquotedblright{}
which corresponds to $\left[-f_{\phi},+f_{\phi}\right]$. In addition,
\emph{for a reasonable construction of the cosmological constant,
we must define the minimum value of the inflaton potential as zero}.
Therefore, the relic of inflaton potential, $V\left(\phi\left(t_{r}\right)\right)>0$,
has survived, only if $\phi\left(t_{r}\right)$ is not at the minimum
position. Returning to \eqref{eq:3.2.7}, the remaining potential
of 
\begin{equation}
V\left(t\right)=\frac{3}{8\pi}\left[H\left(t_{i}\right)-\frac{4\pi}{3}\int_{t_{i}}^{t_{r}}\left(3\dot{\phi}^{2}+4\varepsilon_{\mathrm{r}}\right)d\tau\right]^{2}-\varepsilon_{\mathrm{r}}\left(t_{r}\right)\label{eq:4.2.2-4}
\end{equation}
(where $t\geq t_{r}$) can be obtained. In consequence, the invariable
\eqref{eq:4.2.2-4} plays the role of the energy density of the
ECC $\Lambda$ that appears in the Friedman equations. (The proofs
of $V\left(t\geq t_{r}\right)=V\left(t_{r}\right)$ will be given
later in the \ref{A}ppendix.)
\begin{description}
\item [{Example:}] Consider the case of a classical chaotic model, $V\left(\phi\right)=\frac{1}{2}m_{\phi}^{2}\phi^{2}$
(where $m_{\phi}$ is the mass of inflaton), with $f_{\phi Ms}$ (the
amount of MESFF). The stagnant zone of $\phi$ can be yielded as 
\begin{equation}
\left[\phi_{-},\phi_{+}\right]=\left[-\frac{f_{\phi Ms}}{m_{\phi}^{2}},+\frac{f_{\phi Ms}}{m_{\phi}^{2}}\right].\label{eq:4.2.2-5}
\end{equation}
Obviously, if $\left\Vert \phi\left(t_{r}\right)\right\Vert $ (the
amplitude of $\phi$ at $t_{r}$) is no longer larger than $\frac{f_{\phi Ms}}{m_{\phi}^{2}}$,
the restoring force $V^{'}\left(\phi\left(t_{r}\right)\right)$ will
be cancelled out by the corresponding ESFF, and then $\phi$ will
stop at the position $\phi\left(t_{r}\right)$ forever. The result
is that the remaining inflaton potential has a range of
\begin{equation}
0\leq V\left(\phi\left(t_{r}\right)\right)\leq\frac{f_{\phi Ms}^{2}}{2m_{\phi}^{2}}.\label{eq:4.2.2-6}
\end{equation}
Due to our assumption that $\dot{\phi}\left(t_{i}\right)=0$ when
inflation has just begun, the range of MESFF for our universe can
be found as
\begin{equation}
\sqrt{\frac{m_{\phi}^{2}\Lambda}{4\pi}}\leq f_{\phi Ms}<m_{\phi}^{2}\left\Vert \phi\left(t_{i}\right)\right\Vert ,\label{eq:4.2.2-7}
\end{equation}
where we replace $V\left(\phi\left(t_{r}\right)\right)$ with $\frac{\Lambda}{8\pi}$.
If the amount of EKFF is equal to (or close to) $f_{\phi Ms}$, $f_{\phi Ms}\ll m_{\phi}^{2}\left\Vert \phi\left(t_{i}\right)\right\Vert $
must be obeyed, as otherwise the universe will inflate for a much
longer period.
\end{description}
\end{enumerate}

\section{Conclusions} \label{sec:5}

According to the discussion of the theory of thermo-inflation above,
the \textquotedblleft{}effective friction\textquotedblright{} of a
$\phi$-system will be naturally obtained provided that one accepts
our postulates and Prigogine's suggestion \cite{TC}. Employing the
conclusion of effective friction, we can indicate two special types
of expanding universe: a Type I universe that will enter an expanding
mode with uniformly rolling $\dot{\phi}\left(t_{*}\right)$ after
$t_{*}$; and a Type II universe that will eventually display the
remaining inflaton potential $V\left(\phi\left(t_{r}\right)\right)$,
which plays the role of the ECC $\Lambda$.

In the case of a Type I universe, we obtain the important conclusion
that particles will still be simultaneously created after $t_{*}$.
This is a crucial element in determining the type of universe. Such
a result leads us to propose a toy model for practicing Hoyle's idea.
This is most interesting because several amazing and instructive results
ensue: the Hubble rate is linear with the cosmic time as \eqref{eq:4.2.1-3};
inflaton potential \eqref{eq:4.2.1-5} is equivalent to the form
of $\frac{1}{2}m_{\phi}^{2}\phi^{2}\left(t\right)+V_{0}$; the minimum
potential $V_{0}$ must be negative as in \eqref{eq:4.2.1-8};
and \eqref{eq:4.2.1-9} is consistent with \eqref{eq:3.1.3}.
Moreover, \eqref{eq:4.2.1-12} reveals the fact that a universe
with the condition of \eqref{eq:4.2.1-12} will not expand with
acceleration when it enters the Hoyle course.

For a Type II universe, the invariable \eqref{eq:4.2.2-4} will
be proved in the \ref{A}ppendix. The solution of the remaining potential
is based on the discovery of the stagnant zone of $\phi$ which will
be formed by the $\phi$-system's MESFF. $\phi$ itself will finally
be frozen in this zone with a concluding amplitude of $\left\Vert \phi\left(t_{r}\right)-0\right\Vert $.
Therefore, a nonzero $V\left(\phi\left(t_{r}\right)\right)$ will
survive only if the static position, $\phi\left(t_{r}\right)$, is
not the origin. Additionally, $t_{r}$ must happen before the end
of the radiation-dominated era since the working epoch of our scenario 
is much earlier than the matter-dominated era. Actually, it is very
difficult to state a clear value for $\phi\left(t_{r}\right)$ and
$t_{r}$, not only because the equation \eqref{eq:4.1.2} is highly
nonlinear (the Hubble rate is also the function of EFF), but also
because the quantum probability of the particle creation rate $\Gamma\left(t\right)$
can actually influence the result. It suffices to say that the value
of ECC is probabilistic. Therefore, the conclusion could consistently
explain both the tiny ECC problem and the coincidence problems, even
if $f_{\phi Ms}$ is not close to zero.

Furthermore, the upper limit of the MESFF can be found from the following
condition: it should be smaller than $V^{'}\left(\phi\left(t_{i}\right)\right)$.
If not, $V^{'}\left(\phi\left(t_{i}\right)\right)$ will be cancelled
out by a large MESFF, and lead to the situation of an eternal de Sitter
universe. On the other hand, a universe similar to ours needs a proper
EKFF at the beginning of inflation. Studying \eqref{eq:3.1.6},
\eqref{eq:3.2.6}, \eqref{eq:4.1.2} and \eqref{eq:4.1.2.2}
carefully, we discover that a large EKFF will lead the potential to
keep its value approaching the initial for a long time, and cause
the universe to inflate for a much longer period. Contrarily, an initial
EKFF that is too small will mean that its corresponding rolling $\left\Vert \dot{\phi}\left(t\gtrsim t_{i}\right)\right\Vert $
is not slow enough --- the consumption of inflaton potential in such
a case is fast, which causes the universe to end its inflation early.
Therefore, combining the above discussion and introducing the assumption
of $\left\Vert \mathrm{EKFF}\right\Vert \simeq\left\Vert \mathrm{MESFF}\right\Vert $,
we believe that the size of the stagnant zone will be (very) small
(such as the result of \eqref{eq:4.2.2-5}). This conclusion also
reasonably strengthens our explanation for the tiny ECC problem.

It looks as if a nonzero but very small ECC not only provides an indication
of the existence of inflaton dynamics' effective friction, but also
somewhat explains why our current universe is so huge, but not too
huge.

\acknowledgments{
I sincerely thank Prof. M. Yu. Khlopov, Prof. W.-Y. P. Huang, Dr.
J.-A. Gu and Dr. T.-C. Liu for their useful comments and suggestions;
and the great support received from Marilu Hsu and Vincent Chen. 
I also deeply appreciate my best friend Dan for his important discussion and help. 
Thank my wife Akiko.
Finally, this work was a last gift from my lovely daughter CoCo, and
I thank her for her company during the past 11 years. I hope that
she can receive my thoughts and gratitude to her.
}

\appendix
\section{Proofs} \label{A}

In this appendix, we will provide two proofs of \eqref{eq:4.2.2-4}:
the relic of inflaton potential is
\[
V\left(t\geq t_{r}\right)=\frac{3}{8\pi}\left[H\left(t_{i}\right)-\frac{4\pi}{3}\int_{t_{i}}^{t_{r}}\left(3\dot{\phi}^{2}+4\varepsilon_{\mathrm{r}}\right)d\tau\right]^{2}-\varepsilon_{\mathrm{r}}\left(t_{r}\right)
\]
when $\phi$ itself is finally frozen in the stagnant zone formed
by the MESFF of the $\phi$-system.
\begin{description}
\item [{PROOF 1}]
\end{description}
Because the field $\phi$ must be at rest when $t>t_{r}$, its potential
value is 
\begin{eqnarray}
V\left(t\right) & = & \frac{3}{8\pi}\left[H\left(t_{i}\right)-\frac{4\pi}{3}\int_{t_{i}}^{t}\left(3\dot{\phi}^{2}+4\varepsilon_{\mathrm{r}}\right)d\tau\right]^{2}-\varepsilon_{\mathrm{r}}\left(t\right)\nonumber \\
 & = & \frac{3}{8\pi}\left\{ \left[H\left(t_{i}\right)-\frac{4\pi}{3}\int_{t_{i}}^{t_{r}}\left(3\dot{\phi}^{2}+4\varepsilon_{\mathrm{r}}\right)d\tau\right]^{2}\right.\nonumber \\
 &  & -2\left[H\left(t_{i}\right)-\frac{4\pi}{3}\int_{t_{i}}^{t_{r}}\left(3\dot{\phi}^{2}+4\varepsilon_{\mathrm{r}}\right)d\tau\right]\cdot\left(\frac{4\pi}{3}\int_{t_{r}}^{t}4\varepsilon_{\mathrm{r}}d\eta\right)\nonumber \\
 &  & \left.+\left[\frac{4\pi}{3}\int_{t_{r}}^{t}4\varepsilon_{\mathrm{r}}d\eta\right]^{2}-\frac{8\pi}{3}\varepsilon_{\mathrm{r}}\left(t\right)\right\} ,\label{eq:p.1.1}
\end{eqnarray}
In order to avoid confusion, $\eta$ is employed to replace $\tau$
for the integral-range from $t_{r}$ to $t$. As per \eqref{eq:3.1.2}
and \eqref{eq:3.1.7} with the introduction of \eqref{eq:3.2.4},
the last term of \eqref{eq:p.1.1} is calculated as 
\begin{eqnarray*}
-\frac{8\pi}{3}\varepsilon_{\mathrm{r}}\left(t\right) & = & -\frac{8\pi}{3}\left(\int_{t_{r}}^{t}\dot{\varepsilon}_{\mathrm{r}}d\eta+\varepsilon_{\mathrm{r}}\left(t_{r}\right)\right)\\
 & = & -\frac{32\pi}{9}\int_{t_{r}}^{t}\left(\Gamma\left(t\geq t_{r}\right)-3H\right)\varepsilon_{\mathrm{r}}d\eta-\frac{8\pi}{3}\varepsilon_{\mathrm{r}}\left(t_{r}\right)\\
 & = & \frac{32\pi}{3}\left[H\left(t_{i}\right)-\frac{4\pi}{3}\int_{t_{i}}^{t_{r}}\left(3\dot{\phi}^{2}+4\varepsilon_{\mathrm{r}}\right)d\tau\right]\int_{t_{r}}^{t}\varepsilon_{\mathrm{r}}d\eta\\
 &  & -\frac{512\pi^{2}}{9}\int_{t_{r}}^{t}\left(\int_{t_{r}}^{\eta}\varepsilon_{\mathrm{r}}d\tau\right)\varepsilon_{\mathrm{r}}d\eta-\frac{8\pi}{3}\varepsilon_{\mathrm{r}}\left(t_{r}\right).
\end{eqnarray*}
Further, the above result can be simplified as 
\begin{eqnarray}
-\frac{8\pi}{3}\varepsilon_{\mathrm{r}}\left(t\right) & = & \frac{32\pi}{3}\left[H\left(t_{i}\right)-\frac{4\pi}{3}\int_{t_{i}}^{t_{r}}\left(3\dot{\phi}^{2}+4\varepsilon_{\mathrm{r}}\right)d\tau\right]\int_{t_{r}}^{t}\varepsilon_{\mathrm{r}}d\eta\nonumber \\
 &  & -\frac{256\pi^{2}}{9}\left[\int_{t_{r}}^{t}\varepsilon_{\mathrm{r}}d\eta\right]^{2}-\frac{8\pi}{3}\varepsilon_{\mathrm{r}}\left(t_{r}\right).\label{eq:p.1.2}
\end{eqnarray}
Therefore, taking \eqref{eq:p.1.2} into \eqref{eq:p.1.1}
brings the remaining potential 
\begin{equation}
V\left(t\geq t_{r}\right)=\frac{3}{8\pi}\left[H\left(t_{i}\right)-\frac{4\pi}{3}\int_{t_{i}}^{t_{r}}\left(3\dot{\phi}^{2}+4\varepsilon_{\mathrm{r}}\right)d\tau\right]^{2}-\varepsilon_{\mathrm{r}}\left(t_{r}\right).\label{eq:p.1.3}
\end{equation}

\begin{description}
\item [{PROOF 2}]
\end{description}
For the other, simpler proof, \eqref{eq:3.2.6} and \eqref{eq:3.2.8}
are combined with \eqref{eq:4.1.1}: 
\begin{equation}
\frac{3}{2}\ddot{\phi}\dot{\phi}+\left(3H-\Gamma\right)\dot{\phi}^{2}=\frac{1}{4\pi}\Gamma\dot{H}-\frac{\dot{R}}{32\pi},\label{eq:p.2.1}
\end{equation}
where $R=6\dot{H}+12H^{2}$ is the Ricci curvature. Due to the two
conditions of $\dot{\phi}\left(t\right)$ and $\Gamma\left(t\right)$
being zero when $t\geq t_{r}$, \eqref{eq:p.2.1} must happen
as 
\[
\left.\frac{dR}{dt}\right|_{t\geq t_{r}}=0.
\]
It is equivalent to 
\begin{equation}
\left(6\dot{H}+12H^{2}\right)_{t\geq t_{r}}=C\:\left(\mathrm{constant}\right).\label{eq:p.2.2}
\end{equation}
By bringing \eqref{eq:3.2.1} and \eqref{eq:3.2.2} into \eqref{eq:p.2.2},
we obtain 
\begin{equation}
\left(6\dot{H}+12H^{2}\right)_{t\geq t_{r}}=\frac{16\pi}{3}V\left(t\right)=C.\label{eq:p.2.3}
\end{equation}
According to the time-range of \eqref{eq:p.2.3}, i.e. $t=t_{r}$,
$C$ is equal to $\frac{16\pi}{3}V\left(t_{r}\right)$. 
\begin{equation}
V\left(t\geq t_{r}\right)=V\left(t_{r}\right)=\frac{3}{8\pi}\left[H\left(t_{i}\right)-\frac{4\pi}{3}\int_{t_{i}}^{t_{r}}\left(3\dot{\phi}^{2}+4\varepsilon_{\mathrm{r}}\right)d\tau\right]^{2}-\varepsilon_{\mathrm{r}}\left(t_{r}\right)\label{eq:p.2.4}
\end{equation}
is beyond doubt.

Therefore, according to \eqref{eq:p.1.3} and \eqref{eq:p.2.4},
the ECC is $\Lambda=8\pi V\left(t_{r}\right)$.


\end{document}